\documentclass[twocolumn,10pt,a4paper,float,
amsmath, amssymb, aps,
prl,
]{revtex4-2}

\usepackage[utf8]{inputenc}

\usepackage{float}
\usepackage{graphicx}
\usepackage{dcolumn}
\usepackage{bm}
\usepackage{bbm}
\usepackage{mathrsfs}

\newcommand{\la}{\langle}
\newcommand{\ra}{\rangle}

\DeclareMathOperator{\haf}{haf}

\usepackage[dvipsnames]{xcolor}
\usepackage{hyperref}
\hypersetup{
    colorlinks=true,
    citecolor=BrickRed,
    linkcolor=RoyalBlue
}

\usepackage{ulem}
\newcommand\redout{\bgroup\markoverwith
{\textcolor{red}{\rule[0.5ex]{2pt}{0.8pt}}}\ULon}

\begin{document}

\title{\large Hybrid boson sampling}

\author{V. V. Kocharovsky$^1$\\
\textit{$^{1}$Department of Physics and Astronomy, Texas A\&M University, College Station, TX 77843, USA}}
\date{\today}

\begin{abstract}
We propose boson sampling from a system of coupled photons and Bose-Einstein condensed atoms placed inside a multi-mode cavity as a simulation process testing quantum advantage of quantum systems over classical computers. Consider a two-level atomic transition far-detuned from photon frequency. An atom-photon scattering, and interatomic collisions provide interaction creating quasiparticles and exciting atoms, photons into squeezed entangled states orthogonal, respectively, to the atomic condensate and classical field driving the two-level transition. We find a joint probability distribution of atom and photon numbers within a quasi-equilibrium model via a hafnian of an extended covariance matrix. It shows a sampling statistics that is $\sharp$P-hard for computing even if only photon numbers are sampled. Merging cavity-QED and quantum-gas technologies into hybrid boson sampling setup has the potential to overcome limitations of separate, photon or atom, sampling schemes and reveal quantum advantage.          

\end{abstract}

\maketitle
	
\section{I. Introduction: Overcoming problems of separate, photon or atom, boson sampling by merging the two systems} 

Revealing quantum advantage of many-body quantum systems over classical computers is one of the central themes of modern quantum physics \cite{Harrow2017,Dalzell2020,Zhong2020,Movassagh2023,Aaronson2013}. Since fault-tolerant universal quantum computers equipped with a large-size Hilbert space and quantum error correcting code are out of reach even in the near future, one has to rely on the noisy intermediate-scale quantum computers based on the available or starting-to-emerge technologies \cite{Kaufman2024,Bouland2019,Boixo2018,Arute2019,Castelvecchi2023,Preskill2018,Bremner2017}. Current proposals to reach an intermediate-size asymptotics providing a strong enough evidence for quantum advantage employ sampling problems and specialized quantum simulators that would allow elimination of major dissipation and noise limitation factors \cite{Movassagh2023}. The main sampling schemes are based on boson sampling \cite{Aaronson2013,Kaufman2024}, random circuit sampling \cite{Bouland2019,Boixo2018,Arute2019,Castelvecchi2023} and instantaneous quantum polynomial-time circuits \cite{Bremner2017}. 

{\it Boson sampling} in a linear interferometer fed with photons in specific quantum (Fock, squeezed, etc.) states by external synchronized lasers is the most widely discussed example \cite{Yu2023,Deshpande2022,Bulmer2022,Madsen2022,PanPRL2021,Pan2023,Brod2019,LundPRL2014,Hamilton2019,Quesada2018,Zhong2019,Huh2019,Wang2019}. Recently we suggested \cite{PRA2022,Entropy2022,Entropy2023} {\it atomic boson sampling} from a noncondensed fraction of an equilibrium Bose-Einstein condensed gas as an alternative to the photonic boson sampling. It does not require sophisticated external sources of photons in a prescribed quantum state (due to self-generated squeezing found in \cite{PRA2000}) and eliminate the major limitation factor of boson sampling in a linear interferometer -- an exponential growth of photon losses with increasing number of channels taking place due to an inevitable increase in the number of intermode couplers (beam splitters, phase shifters, etc.) needed for coupling each input channel with every output channel. Yet, it requires a multi-detector system measuring occupation numbers of a set of orthogonal excited atom states with a single-atom resolution and close to 100 percent efficiency, which is not available yet. 

The aforementioned and some other problems of the separate photon and atom samplings precluded reaching large-size asymptotics and enough clearness in boson sampling experiments for definitive demonstration of quantum advantage, although the results of recent experiments on Gaussian boson sampling of photons in the 216- and 144-mode interferometers \cite{Madsen2022,PanPRL2021} and ultracold atoms in a tunnel-coupled optical lattice \cite{Kaufman2024} were truly remarkable. 

Here we propose hybrid boson sampling from a coupled atom-photon many-body system combining advantages of two state-of-the-art, quantum-gas and cavity-QED, technologies. It allows one to eliminate sophisticated sources of squeezed photons and exponentially scaling photon losses in the linear interferometer as well as simultaneously solve the problem of multi-detector atom number measurement by using well-developed photon detectors. Measuring numbers of photons alone is already enough for revealing quantum advantage. Yet, with emergence of the detectors for atom numbers, the combination of the BEC-gas and QED-cavity sampling setups could become an ultimate stage for studying quantum advantage. 

The system consists of a Bose-Einstein-condensed, quasi-equilibrium weakly-interacting gas of $N$ two-level atoms placed inside a multi-mode cavity and pumped by a coherent classical laser field. The frequencies of all optical fields are far-detuned from the two-level atomic transition. So, the atom-photon scattering is elastic and does not destroy Bose-Einstein condensate (BEC) by an excessive heating through spontaneous emission since the upper level population is negligibly small.

Such setups had been successfully implemented experimentally back in 2007 in Berkeley \cite{Stamper-Kurn2007}, Zürich \cite{Brennecke2007}, Tübingen \cite{Slama2007}, and Paris \cite{Colombe2007}. However, since then the studies of such systems (see reviews \cite{Ritsch2021,Kirton2019,Mekhov2012,Esslinger2013} and references therein) were mainly focused on the modeling various condensed-matter Hamiltonians (Bose-Hubbard, Ising, Heisenberg, Dicke, etc.) and corresponding phase transitions, associated with mean-field restructuring of the system to Mott insulator, quasicrystal, superradiant and alike phases, as well as on other applications such as laser cooling of quantum gases \cite{Esslinger2013,Wolke2012} or their non-demolition measurements. The analysis of quantum fluctuations around the mean-field values was usually restricted to the studies of just second order correlations. So, the analysis of the $\sharp$P-hard computational complexity of quantum many-body statistics of such systems, which requires a full evaluation of a joint probability distribution of various quantum quantities, i.e., moments or cumulants of all higher orders, has been missing until now.  

In essence, the idea is to employ a quantum BEC gas as a nonlinear optical element inside a multi-mode cavity for producing squeezed entangled states of atoms, photons. The interacting BEC gas not only replaces the lossy intermode couplers and sophisticated external photon sources based on the on-demand parametric oscillators, but also introduces, in addition to quantum two-level (qubit) internal atomic degrees of freedom, the quantum atomic degrees of freedom associated with the translational motion of atoms. As a result, one gets a versatile fine-tunable profoundly quantum interacting many-body system perfectly suitable for examining quantum advantage.       
 
We calculate (within a quasi-equilibrium model) the characteristic function and joint probability distribution of atom numbers (for any set of bare-atom excited states) and photon numbers (for any preselected set of modes) via the covariance matrix.
It depends on the interaction and pump laser parameters, geometry of the system and unitary transformations between the basis of excited atom states and photon modes chosen for sampling and the bases of atom-photon quasiparticles and eigen-squeeze modes. 
As a result, in virtue of the hafnian master theorem \cite{LAA2022} and the fact that computing the hafnian in a general case is $\sharp$P-complete \cite{Barvinok2016}, the statistics of such a mixed (atom-photon) boson sampling turns out to be $\sharp$P-hard for computing. This fact implies that quantum advantage manifestations should be observed.

\section{Multi-mode cavity QED for BEC gas of two-level atoms coupled to photons}

Let us consider Bose-Einstein condensation and related low-temperature/energy cavity-QED phenomena in a dilute weakly interacting gas of spinless Bose atoms having an optical transition of a frequency $\omega_a$ and dipole moments ${\bf d}_a$. Within the second-quantization representation of the nonrelativistic quantum field theory \cite{FetterWalecka}, such a many-body system of identical particles is described by two annihilation field operators $\hat{\psi}_1({\bf r}), \hat{\psi}_2({\bf r})$ acting in a symmetrized Hilbert space. They describe quantum behavior of two-level atoms, occupying the 1-st (lower) or 2-nd (upper) levels, respectively, in regard to the position ${\bf r}$ in space, that is, the translational degree of freedom.   

The gas is kept inside a multi-mode cavity by a classical, say, magneto-optical, trapping potential $V_{\rm ext}({\bf r})$ and is driven by a laser with a classical coherent electrical field of a complex amplitude $E_0({\bf r})$, polarization vector ${\bf e_0}$ and frequency $\omega_0$. The energy of its interaction with an atom is described by Rabi frequency $\Omega_0({\bf r}) = {\bf d}_a{\bf e}_0E_0/\hbar$. The cavity supports a set of $M_{\rm{ph}}$ high-Q modes with an electrical field of complex amplitude ${\bf e}_{\nu}E_{\nu}({\bf r}), \nu = 1,...,M_{\rm{ph}}$, polarization vector ${\bf e_{\nu}}$, frequency $\omega_{\nu}$. Cavity QED of these Bose modes employs their annihilation operators $\{ \hat{b}_{\nu} \}$ acting in the Fock space. 

The frequencies of all fields are far detuned from the atomic transition frequency: $\Delta_a \equiv \omega_a - \omega_0, \omega_a - \omega_{\nu} \gg \gamma$, where $\gamma = T_2^{-1}$ is the decay rate of the atomic dipole. In this limit the upper level population is negligibly small and the upper-level field operator $\hat{\psi}_2({\bf r})$ can be adiabatically eliminated from the Heisenberg equations, so that the many-body system of $N$ trapped atoms interacting with $M_{ph}$ modes in the high-finesse optical cavity is described by a well-known Hamiltonian \cite{Ritsch2021} 
\begin{equation} \label {H}
\begin{split}
&\hat{H} = \sum_{\nu} \hbar\Delta_{\nu}\hat{b}_{\nu}^\dagger \hat{b}_{\nu} + 
\int \hat{\psi}_a^\dagger \left[ \hat{H}_a + \hat{H}_{a-a} + \hat{H}_{a-ph} \right] \hat{\psi}_a d^3{\bf r}, \\
&\hat{H}_a = -\frac{\hbar^2}{2m}\nabla^2 +V_{{\rm ext}}({\bf r}) + \frac{\hbar |\Omega_0({\bf r})|^2}{\Delta_a} , 
\ \hat{H}_{a-a} = \frac{g_a}{2} \hat{\psi}_a^\dagger \hat{\psi}_a,\\ 
&\hat{H}_{a-ph} = \frac{\hbar}{\Delta_a}\sum_{\nu} \Big[ \Omega_{\nu}^*({\bf r})\Omega_0({\bf r})\hat{b}_{\nu}^\dagger + \Omega_{\nu}({\bf r})\Omega_0^*({\bf r})\hat{b}_{\nu} \Big]\\ 
&\qquad \qquad +\frac{\hbar}{\Delta_a}\sum_{\nu,\nu'} \Omega_{\nu}^*({\bf r})\Omega_{\nu'} ({\bf r}) \hat{b}_{\nu}^\dagger \hat{b}_{\nu'} .
\end{split}
\end{equation}
It is written in the frame rotating with the frequency of the classical driving field. So, 
the first term, representing energies of the bare cavity modes, $\hbar \omega_{\nu} \hat{q}_{\nu}$, involves detunings $\Delta_{\nu} = \omega_{\nu} - \omega_0$. The operator $\hat{q}_{\nu} = \hat{b}_{\nu}^\dagger \hat{b}_{\nu}$ gives the number of quanta in a bare cavity mode $\nu$. $\hat{H}_a$ is an effective single-atom Hamiltonian accounting for two trap potential: the external one, $V_{\rm ext}({\bf r})$, and the one created by the far-off-resonance classical field, $\hbar |\Omega({\bf r})|^2/\Delta_a$. The term $\hat{H}_{a-a}$ is responsible for the interatomic interaction determined by the $s$-wave scattering length $a_a$ via the parameter $g_a = 4\pi a_a \hbar^2/m$, where $m$ is an atom mass. The last term $\hat{H}_{a-ph}$ described the atom-photon interaction via (i) creation or annihilation of a photon in the $\nu$-th cavity mode due to scattering on atoms from or into the classical driving mode and (ii) photon exchange between the $\nu$-th and $\nu'$-th modes mediated by scattering on atoms. Hereinafter the lower-level atom field operator is denoted as $\hat{\psi}_a \equiv \hat{\psi}_1({\bf r}) = \sum_l \phi_l({\bf r})\hat{a}_l$. The factor $\Omega_{\nu}({\bf r}) = {\bf d}_a{\bf e_{\nu}}E_{\nu}/\hbar$ is the single-photon Rabi frequency determined by the electrical field ${\bf e_{\nu}}E_{\nu}({\bf r})$ of the $\nu$-th mode. The field profile is normalized in such a way that the electromagnetic energy density integrated over the volume occupied by the cavity mode is equal to the energy of a single photon, $\int |E_{\nu}|^2 d^3{\bf r}/(2\pi) = \hbar \omega_{\nu}$. 

If a weak relaxation and dissipation of both photon and atom bosons described by annihilation operators $\{ \hat{c}_j \} = \{ \{ 
\hat{a}_l\}, \{\hat{b}_{\nu} \} \}$ is important, it can be accounted for in a Born-Markov-RWA approximation by Lindblad equation for the atom-light density operator via decay rates $2\kappa_j$, 
\begin{equation} \label{rho-Eq}
\begin{split}
\frac{d\hat{\rho}}{dt} = &-\frac{i}{\hbar} [\hat{H}, \hat{\rho}] 
+\sum_j \kappa_j \bar{n}_j (2\hat{c}_j^\dagger \hat{\rho} \hat{c}_j - \hat{c}_j \hat{c}_j^\dagger \hat{\rho} - \hat{\rho} \hat{c}_j \hat{c}_j^\dagger ) \\
&+\sum_j \kappa_j (1+\bar{n}_j) (2\hat{c}_j \hat{\rho} \hat{c}_j^\dagger - \hat{c}_j^\dagger \hat{c}_j \hat{\rho} - \hat{\rho} \hat{c}_j^\dagger \hat{c}_j ).
\end{split}
\end{equation}
Here $\bar{n}_j$ is a thermal population of a bath's mode resonantly coupled to a partial boson mode $j$. For simplicity's sake, it is written in the case of independently decaying modes, without cross-mode coupling $\hat{c}_j\hat{c}_{j'}^\dagger$ via a bath \cite{Jager2022}. 

Importantly, an interaction (scattering) between atoms and photons is strongly enhanced for the specially designed high-Q modes since the photons, before leaking the cavity, traverse atom cloud a huge number of times, $Q \ggg 1$, being reflected by cavity mirrors. For low-Q modes, an interaction between atoms and photons is greatly suppressed and their population is negligible. As a result, the low-Q modes are excluded from Eqs.~(\ref{H})-(\ref{rho-Eq}). 

In general, the above system is an open, dissipative driven system that, after placing an equilibrium (at temperature $T_0$) BEC gas inside an initially empty (no photons) optical cavity, evolves towards some steady state with nonzero photon occupations in virtue of the pump laser light scattering on atoms. In some cases \cite{Ritsch2021,Lukin2013,Gorshkov2016,Lebreuilly2018}, this state may be approximated as a quasi-equilibrium state with some effective temperature $T$ which accommodates the effects of the initial gas temperature $T_0$, leakage of atoms from the trap (in particular, due to three-body collisions, trap's imperfectness), duration, intensity and noise of the laser pump, cavity-loss-induced noise, etc.

\section{Eigen-squeeze modes $\&$ quasiparticles vs. excited bare atoms $\&$ photons}

The aforementioned quasi-equilibrium state is favored once the atom-photon scattering is strong but the losses of photons and atoms are very low, so that the system evolves longer than a characteristic scattering time which is estimated \cite{Ritsch2021,Piazza2014} as $\tau_{\rm s} \sim N\kappa_{\nu}^3\Delta_a^2/(\Delta_{\nu}\Omega_0^2\Omega_{\nu}^2\omega_r)$; $\omega_r = \frac{\hbar\omega_{\nu}^2}{2mc^2}$ is the recoil frequency. In this case atoms and photons, which constitute supermode polaritons \cite{Ritsch2021}, form hybrid atom--photon quasiparticles and have enough time to equilibrate. In particular, the cavity photons cool or heat atoms \cite{Esslinger2013,Wolke2012,Piazza2014} towards a thermal state with temperature $T \sim \hbar \kappa_{\nu}$ if $|\Delta_{\nu}| \gg \omega_r$. Short-range collisions between atoms also benefit a thermal steady state \cite{Bezvershenko2020}.

Let us model a system state by a quasi-equilibrium density operator $\hat{\rho} = e^{-\hat{H}_{\rm eff}/T}/{\rm Tr}\{ e^{-\hat{H}_{\rm eff}/T} \}$ (see \cite{Ritsch2021,Lukin2013,Gorshkov2016,Lebreuilly2018}) which represents a possible quantum statistics of the relevant atom and photon modes via an effective quadratic Hamiltonian $\hat{H}_{\rm eff}$. In general, such a Gaussian state is more classical and mixed than other, more pure quantum states. So, if its boson-sampling statistics is $\sharp$P-hard for computing, than boson sampling in other dynamical non-equilibrium or steady quantum states is even more prone to $\sharp$P-hardness. 
Such states will be discussed elsewhere. 
Here we just note that squeezing required for the $\sharp$P-hardness is generated via non-equilibrium processes both in the photon and atom modes \cite{Piazza2014}. 

In the limit of very weak losses the coupled atoms and photons, both obeying the Bose statistics, tend to form some kind of a Bose-Einstein-condensed gas. If cavity supports BEC of photons (like in photon BEC \cite{Schmitt2014,Wang2019photonBEC}, when photon reabsorption via rovibrational dye manifold in an intracavity reservoir/bath dominates over photon losses), then, even after switching off the pump laser, quasi-equilibrium macroscopic condensates for both atom and photon components could be formed. In any case, we skip discussion of atom and photon condensates, described by equations similar to the Gross-Pitaevskii one, and denote the related classical fields as $\phi_0({\bf r})$ and $\Omega_0({\bf r})$. 

One can think of the optical driving field, $E_0({\bf r})$, or its Rabi frequency, $\Omega_0({\bf r})$, as a kind of photon condensate if a coherent scattering of the drive on the atom condensate due to linear in photon operators $\hat{b}_{\nu}^\dagger, \hat{b}_{\nu}$ terms in Eq.~(\ref{H}) is set aside \cite{Jager2022}. Both the photon condensate and drive laser field are macroscopic coherent fields scattering from which (or, in the words adopted in BEC physics, quantum depletion of which) populates the noncondensed high-Q cavity modes with photons, on top of the aforementioned coherent component if any. One can infer from Eq.~(\ref{H}) a model Hamiltonian $\hat{H}_{\rm{eff}}$, describing the statistical operator $\hat{\rho}$ of the quasi-equlibrium, BEC-like phase of the hybrid atom-photon quasiparticles. 

Following Bogoliubov-Popov approach \cite{Shi1998}, we replace the operator annihilating photon in the mode $E_0({\bf r})$ by a c-number, $\hat{b}_0 \approx \sqrt{q_0}$, assuming that a mean number of quanta (photons) is large, $q_0 \gg 1$. So, the photon field operator is $\hat{\psi}_{\rm ph}({\bf r}) = \mathcal{E}_0({\bf r})\sqrt{q_0} + \sum_{\nu \neq 0}\mathcal{E}_{\nu}({\bf r}) \hat{b}_{\nu}$, where $\mathcal{E}_{\nu} = E_{\nu}({\bf r})/[\int |E_{\nu}|^2 d^3{\bf r}]^{1/2}$.
Similarly, we approximate the atom field operator by a sum of its classical part and small quantum excitations,
$\hat{\psi}_a({\bf r}) = \phi_0({\bf r})\sqrt{N_0} + \sum_{l\neq 0} \phi_l({\bf r})\hat{a}_l$, where $N_0$ is a mean number of condensed atoms and $\hat{a}_l, l \neq 0,$ is an operator annihilating an atom in a bare-atom excited state $\phi_l$ orthogonal to $\phi_0$. All wave functions are normalized to unity, $\int |\phi_l|^2 d^3{\bf r} = 1$. Keeping in (\ref{H}) only terms quadratic in operators $\hat{a}_l$, $\hat{b}_{\nu}$, we get the effective Hamiltonian of Bogoliubov-Popov type
\begin{equation} \label{BH}
\begin{split}
&\hat{H}_{\rm{eff}} = \frac{1}{2}\left( \begin{matrix} \hat{{\bf c}}^\dagger \\ \hat{{\bf c}} \end{matrix} \right)^T \! H \! \left(  \begin{matrix} \hat{{\bf c}}^\dagger \\ 
\hat{{\bf c}} \end{matrix} \right), 
\qquad H =\left[ \begin{matrix} \tilde{\chi}  
&   \epsilon + \chi\\ \epsilon + \chi^*   
&  \tilde{\chi}^*     \end{matrix} \right]; \\
&\epsilon = \left[ \begin{matrix} \epsilon_a  
&   0 \\ 0   
& \epsilon_{\rm ph} \end{matrix} \right], \qquad \epsilon_{\rm ph} = {\rm diag} \{\hbar\omega_{\nu} \}, \\
&\epsilon_a = \left( \int \phi_l^* \left[ \hat{H}_a -\mu +2g_a(N_0 |\phi_0|^2 + n_{\rm ex}) \right] \phi_{l'} \, d^3{\bf r} \right), \\
&\chi = \left[ \begin{matrix} 0 
& \chi_{\rm a-ph} \\ \chi_{\rm ph-a} 
& \chi_{\rm ph-ph} \end{matrix} \right], \qquad \tilde{\chi} = \left[ \begin{matrix} \tilde{\chi}_{a-a} 
& \tilde{\chi}_{\rm a-ph} \\ \tilde{\chi}_{\rm ph-a}   
& 0 \end{matrix} \right].
\end{split}
\end{equation}
It is a quadratic form in the creation, $\hat{{\bf c}}^\dagger = \{ \{ \hat{a}_l^\dagger\}, \{\hat{b}_{\nu}^\dagger \} \}^T$, and annihilation, $\hat{{\bf c}} = \{ \{ 
\hat{a}_l\}, \{\hat{b}_{\nu} \} \}^T$, 2-block column vector operators combining the atom and photon operators. The superscript $T$ denotes a transpose of a vector or matrix. The form's $(2\times 2)$-block $2M\times 2M$ matrix $H$ is built of diagonal ($\chi, \tilde{\chi}$) and off-diagonal ($\epsilon + \chi, \epsilon + \chi^*$) square blocks of size $M\times M$, where $M=M_a+M_{ph}$ with $M_a$ and $M_{ph}$ being, respectively, the numbers of bare-atom excited states $\{\phi_l | l=1,\ldots, M_a\}$ and high-Q cavity modes $\{E_{\nu} | \nu=1,\ldots, M_{ph}\}$ which notably contribute to the state of the atom-photon system. The star $^*$ denotes a complex conjugate, $\mu$ is a chemical potential, $n_{\rm ex}({\bf r})$ a mean density of the noncondensate. The block $\epsilon$ itself is a $(2\times 2)$-block matrix --  a diagonal matrix built of the $M_a\times M_a$ matrix $\epsilon_a$ and $M_{\rm ph}\times M_{\rm ph}$ matrix $\epsilon_{\rm ph}$ which originate from the single-atom, $\hat{H}_a$, and single-mode, $\hbar\omega_{\nu}\hat{b}_{\nu}^\dagger\hat{b}_{\nu}$, energy contributions in Eq.~(\ref{H}), respectively. The blocks $\chi$, $\tilde{\chi}$ themselves are also $(2\times 2)$-block matrices. They constitute an analogue of the matrix of Bogoliubov couplings between bare-atom excited states and high-Q photon modes and cross-couplings:
$\tilde{\chi}_{a-a} = \left( g_aN_0 \int \phi_l^* \phi_{l'}^* \, \phi_0^2 \, d^3{\bf r} \right),$ 
$\chi_{\rm ph-ph} = \left( \frac{\hbar N_0}{\Delta_a} \int \Omega_{\nu}^*\Omega_{\nu'} |\phi_0|^2 \, d^3{\bf r} \right),$
$\chi_{\rm ph-a} = \left( \frac{\hbar\sqrt{N_0}}{\Delta_a}\int \Omega_{\nu}^*\Omega_0 \phi_{l'}\phi_0^* \, d^3{\bf r} \right), \ \chi_{\rm a-ph} = \chi_{\rm ph-a}^\dagger,$
$\tilde{\chi}_{\rm ph-a} = \left( \frac{\hbar\sqrt{N_0}}{\Delta_a}\int \Omega_{\nu}^*\Omega_0 \phi_{l'}^*\phi_0 \, d^3{\bf r} \right), \ \tilde{\chi}_{\rm a-ph} = \tilde{\chi}_{\rm ph-a}^\dagger.$

The principal part in the quantum advantage and $\sharp$P-hardness of the above many-body system is played by the matrix $\tilde{\chi}$ which bears the counter-rotating (cf. non-RWA, beyond the rotation wave approximation) atom-atom $(\tilde{\chi}_{a-a})_{ll'} \hat{a}_l^\dagger \hat{a}_{l'}^\dagger$ and photon-atom $(\tilde{\chi}_{\rm ph-a})_{\nu l'} \hat{b}_{\nu}^\dagger \hat{a}_{l'}^\dagger$ couplings. (An off-resonance optical response of two-level atoms in the ground state does not include appreciable photon-photon counter-rotating terms.) The matrix $\chi$ bears the usual co-rotating (cf. RWA) atom-photon $(\chi_{a-ph})_{l\nu'} \hat{a}_l^\dagger \hat{b}_{\nu'}$, photon-atom $(\chi_{ph-a})_{\nu l'} \hat{b}_{\nu}^\dagger \hat{a}_{l'}$ and photon-photon $(\chi_{\rm ph-ph})_{\nu \nu'} \hat{b}_{\nu}^\dagger \hat{b}_{\nu'}$ couplings.
The atom--atom coupling block $\tilde{\chi}_{a-a}$ is a square $M_a\times M_a$ matrix, while a photon-photon coupling block $\chi_{\rm ph-ph}$ is a square $M_{\rm ph}\times M_{\rm ph}$ matrix. The photon-atom and atom-photon blocks $\chi_{\rm ph-a}$, $\chi_{\rm a-ph}$ and $\tilde{\chi}_{\rm ph-a}$, $\tilde{\chi}_{\rm a-ph}$ are Hermitian conjugated rectangular $M_{\rm ph}\times M_a$ and $M_a\times M_{\rm ph}$ matrices. 

With the help of the effective Hamiltonian (\ref{BH}) derived above, we can solve the problem on quantum statistics of the mixed atom-photon sampling by generalizing the method which has been developed in \cite{PRA2022,Entropy2022,Entropy2023} for the pure atom sampling from BEC gas. The crucial point of this method is finding the coupled atom-photon eigen-squeeze modes along with the eigen-energy quasiparticles. Note that the eigen-squeeze modes are uniquely defined for the many-body interacting system and are as important for its quantum many-body statistics as the quasiparticles for the mean-field, thermodynamic characteristics. In particular, an existence of the eigen-squeeze modes with relatively large eigenvalues (i.e., single-mode squeezing parameters) is required for the emergence of the computational $\sharp$P-hardness and quantum advantage. 

We find the solution via the irreducible Bloch-Messiah reduction \cite{Braunstein2005,BlochMessiahPRA2016,Vogel2006,Huh2017} of Bogoliubov transformation $\tilde{R}$ from the bare operators to quasiparticle operators $\hat{\tilde{{\bf c}}}^\dagger, \hat{\tilde{{\bf c}}}$. It is
\begin{equation} \label{BlochMessiah}
\begin{split}
&\tilde{R} = \tilde{R}_W \tilde{R}_r \tilde{R}_V,  
\quad \bigg(  \begin{matrix}  {\,\bf \hat{\tilde{c}}^\dagger}   \\
                            {\bf \hat{\tilde{c}}}           \end{matrix} \bigg)
= \tilde{R} \bigg(  \begin{matrix}  {\,\bf \hat{c}^\dagger}   \\
{\bf \hat{c}}           \end{matrix} \bigg); \quad \tilde{R}_V = \bigg[  \begin{matrix}  V^*   &   0   \\
                            0   &   V \end{matrix} \bigg], \\
&\tilde{R}_W = \bigg[  \begin{matrix}   W^*     &   0       \\
                                        0       &   W     \end{matrix} \bigg], \ 
\tilde{R}_r = \bigg[  \begin{matrix}  \cosh\,\Lambda_r    &   \sinh\,\Lambda_r    \\
                            \sinh\,\Lambda_r    &   \cosh\,\Lambda_r    \end{matrix} \bigg]. 
\end{split}    
\end{equation}
It follows from a singular value decomposition of the blocks of the Bogoliubov-transformation matrix:
\begin{equation} \label{Bblocks}
\tilde{R} = \bigg[ \begin{matrix} A^* & -B^* \\ -B & A        \end{matrix} \bigg]; 
A = W \cosh \Lambda_r V, \ B = -W \sinh \Lambda_r V^* .
\end{equation}

The $M\times M$ unitary matrix $V$ describes a transformation between operators annihilating excitations in the bare states, $\{\hat{c}_j\}$, and in the eigen-squeeze modes, $\{\hat{\beta}_j\}$. It is equivalent to a basis rotation in the single-particle Hilbert space from the bare basis of atom and photon excited, noncondensate states $\{ \phi_j| j=1,...,M_a\} \cup \{ \phi_j = \mathcal{E}_{j-M_a}| j=M_a+1,...,M \}$ to the basis of coupled atom-photon eigen-squeeze modes $\{ \varphi_j, j=1,\ldots,M \}$, that is,
\begin{equation} \label{unitaryMixingV}
\hat\beta_j = \sum_{j'=1}^M V_{jj'} \hat{c}_{j'}, \
\varphi_j = \sum_{j'=1}^M V^*_{jj'} \phi_{j'},
\ \hat{\psi}_{\rm ex}({\bf r}) = \sum_{j=1}^M \varphi_j({\bf r}) \hat\beta_j .
\end{equation}
A field operator $\hat{\psi}_{\rm ex}({\bf r})$ in Eq.~(\ref{unitaryMixingV}) combines partial, bare atom and photon field operators $\hat{\psi}_a({\bf r})$ and $\hat{\psi}_{\rm ph}({\bf r})$. It annihilates a quantum of the coupled atom-photon excitations in the eigen-squeeze modes (not quasiparticles). 

The central part, $\tilde{R}_r$, of the Bloch-Messiah reduction is not an identity matrix due to the counter-rotating terms. It upgrades the atom-photon field operator to the form,
\begin{equation} \label{r}
\hat{\psi}_{\rm ex} = \sum_{j=1}^M ( u_j' \hat{c}_j' + v_j'^* \hat{c}_j'^\dagger ), 
u_j' = \varphi_j \cosh r_j, v_j'^* = -\varphi_j \sinh r_j,
\end{equation}
mixing annihilation and creation operators of the eigen-squeeze modes, $\hat{c}_j' = \hat{\beta}_j \cosh r_j + \hat{\beta}_j^\dagger \sinh r_j$. It sets a two-component functional space with a basis $\{ u_j'({\bf r}), v_j'^*({\bf r}) \}$ defining two-component eigen-squeeze excitations characterized by a single-mode squeezing parameter $r_j \geq 0$. They are the eigenvalues of a multimode squeeze matrix $r = W\Lambda_rW^\dagger$ and constitute the matrix $\Lambda_r = {\rm diag} \{ r_j \}$. 

The $M\times M$ unitary $W$ converts operators of the two-component eigen-squeeze excitations into polariton operators $\hat{\tilde{c}}_j$ diagonalizing Hamiltonian: $\hat{H}_{\rm{eff}} = \sum_j \tilde{E}_j \hat{\tilde{c}}_j^\dagger \hat{\tilde{c}}_j$, 
\begin{equation} \label{BW}
\hat{\tilde{c}}_j = \sum_{j'=1}^M W_{jj'} \hat{c}_{j'}', \
\ \hat{\psi}_{\rm ex}({\bf r}) = \sum_{j=1}^M ( u_j({\bf r}) \hat{\tilde{c}}_j + v_j^*({\bf r}) \hat{\tilde{c}}_j^\dagger );
\end{equation}
$u_j = \sum_{j'} W_{jj'}^* \varphi_{j'} \cosh r_{j'}, \ v_j^* = -\sum_{j'} W_{jj'} \varphi_{j'} \sinh r_{j'}$.

\section{Quantum statistics of hybrid photon-atom sampling via hafnian master theorem} 

Once the matrix of Bogoliubov transformation is calculated, we find the $2M\times 2M$ covariance matrix of the atom-atom, photon-photon and atom-photon correlators:
\begin{equation} \label{G}
G \equiv \left[ \begin{matrix}
\big(\la \hat{c}_j^\dagger \hat{c}_{j'} \ra\big) 
            &   
\big(\la \hat{c}_j^\dagger \hat{c}_{j'}^\dagger \ra\big)
            \\[6pt]
\big(\la \hat{c}_j \hat{c}_{j'} \ra\big) 
            &   
\big(\la \hat{c}_j^\dagger \hat{c}_{j'} \ra\big)
        \end{matrix} \right] = \frac{1}{2} R
\begin{bmatrix}  Q  &  0 
                            \\
 0  &  Q   \end{bmatrix}  R^\dagger - \frac{1}{2},        
\end{equation}
where $Q = \textrm{diag} \{\coth \frac{\tilde{E}_j}{2T}|j=1,...,M \}$ and $R = \tilde{R}^{-1}$. 

Finally, applying the method of the characteristic function developed in \cite{PRA2022,Entropy2022,Entropy2023} and the hafnian master theorem \cite{LAA2022}, we find the joint probability distribution of atom and photon numbers $\{ \{N_l|l=1,...,M_a\}, \{q_{\nu}|\nu = 1,...,M_{\rm ph} \} \}$ sampled by a simultaneous multi-detector measurement over a set of $M_a$ excited-atom states and $M_{\rm ph}$ cavity modes selected from the noncondensate ones: 
\begin{equation} \label{pdf=Hafnian}
\rho\big(\{ \{N_l\}, \{ q_{\nu}\} \}\big) = \frac{ \haf \tilde{C}(\{ \{N_l\}, \{ q_{\nu}\} \})}{\sqrt{\det (1+G)} (\prod_l N_l!) \prod_{\nu} q_{\nu}!} \ . 
\end{equation}

It is given by the hafnian of the $(2n\times 2n)$ extended covariance-related matrix $\tilde{C}$, where $n = \sum_l N_l + \sum_{\nu} q_{\nu}$ is the total number of counts in a sample for all detector channels, including all excited-atom states $\{ \phi_l \}$ and photon modes $\{ \mathcal{E}_{\nu}\}$ chosen for sampling. The matrix $\tilde{C}$ is a certain extension of a covariance-related matrix $C = PG(1 +G)^{-1}$. Namely, the $\tilde{C}$'s $l$-th and $(M+l)$-th rows are replaced with $N_l$ copies of the $l$-th and $(M+l)$-th rows, accordingly. Then, $l$-th and $(M+l)$-th columns are replaced with $N_l$ copies of the $l$-th and $(M+l)$-th columns. Finally, a similar replacement is done with $(M_a+\nu)$-th and $(M+M_a+\nu)$-th rows as well as with $(M_a+\nu)$-th and $(M+M_a+\nu)$-th columns using $q_{\nu}$ their copies. 
The matrix $P$ permutes the off-diagonal and diagonal blocks of the $(2\times 2)$-block matrix $G(1+G)^{-1}$.

\section{Multi-detector measurements for sampling photon and atom numbers} 

The challenge of photon-atom sampling experiments is in simultaneous measurement of photon numbers $\{q_{\nu}|\nu = 1,...,M_{\rm ph} \}$ and atom numbers $\{N_l|l=1,...,M_a\}$ in the noncondensate optical cavity modes and atom-excited states with a single-photon/atom resolution. Moreover, parameters of the BEC-gas $\&$ QED-cavity setup, including the number of trapped atoms, temperature, BEC trap and multi-mode cavity geometries, their mutual alignment, parameters of the pump laser and so on, should be precisely controlled and identified or post-selected.   

Such measurements could be based on a nondestructive multi-detector imaging of atoms in each of $M_a$ excited states and a nondemolishing monitoring of photon numbers in high-Q cavity modes via detecting photons escaping each of $M_{\rm ph}$ modes. A destructive measurement, say, by quenching the BEC trap potential and making transparent the optical cavity, is another possibility. 

A required technique for multi-mode photon counting is already available in quantum optics. Measuring and sampling atom number fluctuations in the noncondensed fraction of a BEC gas is coming soon as is evident from promising works related to this problem \cite{Kaufman2024,Kaufman2018,Kristensen2019,Christensen2021,Clement2021,Clement2023,Robens2024,Pit2011,Castin,Kristensen2017,Jacqmin2010,Armijo2010,RaizenBECstatisticsPRL2005,AspectDensityFluctPRL2006,Dotsenko2005,Schlosser2002,Raizen2009}. 
A successful experiment on measuring fluctuations in the total number of noncondensed atoms has been reported in \cite{Kristensen2019,Christensen2021}. Thus, the main difficulty of such measurements -- a differentiation of the noncondensate from much more populated condensate \cite{PRA2020} -- has been resolved.  

A striking time of flight experiments on recording atom numbers in various momentum states of a BEC gas based on the position the atom impacts on a detector array after a free fall of the atom cloud due to gravity have been done in \cite{Clement2021,Clement2023}. Their detectors showed a single atom resolution. 
A boson sampling machine with atoms has been shown in \cite{Robens2024} by revealing the Hong-Ou-Mandel interference of two Bose atoms in a 4-mode interferometer. 

Importantly, the results in Eqs.~(\ref{G})-(\ref{pdf=Hafnian}) show that for unveiling manifestations of $\sharp$P-hardness and quantum advantage it is enough to detect just photon numbers. A cavity-QED technique for such a sampling is readily available and could be similar to photon BEC technique \cite{Schmitt2014,Wang2019photonBEC}. So, even using BEC gas only as a nonlinear optical element producing squeezed states, that is, not including atom-number detector channels into a sampling ensemble ($M_a =0$), we still get a very general form of the covariance matrix $G$ generating the extended covariance matrix $\tilde{G}$, hafnian of which in Eq.~(\ref{pdf=Hafnian}) is $\sharp$P-hard for computing. The point is that the photon-atom coupling (\ref{BH}) results in the atom-photon entanglement and generates squeezing and complexity of photon states of the high-Q cavity modes (supermode polaritons \cite{Ritsch2021}) due to the symplectic Bogoliubov transform (\ref{BlochMessiah}), (\ref{Bblocks}), similar to that happening for the pure atomic boson sampling due to atom-atom coupling in a BEC gas alone \cite{PRA2022,Entropy2022,Entropy2023}. 

In fact, the result in Eqs.~(\ref{BlochMessiah}), (\ref{Bblocks}) means that the BEC gas in a QED cavity possesses two intrinsic, naturally built-in interferometers linked to the unitaries $V$ and $W$. In the case of just photon sampling ($M_{\rm ph}\neq 0, M_a = 0$), they are $M_{\rm ph}\times M_{\rm ph}$ matrices whose $M_{\rm ph}^2$ entries could be arbitrarily varied due to a functional freedom in choosing (a) the sampling modes selected for detecting and (b) the trapping potential. Obviously, this is equivalent to having a random Gaussian unitary inside the matrix $\tilde{G}$ under the hafnian in Eq.~(\ref{pdf=Hafnian}) with $\sim M_{\rm ph}^2$ independently variable parameters and no degeneracy. (Eq.~(\ref{G}) just adds an extra mixing.) So, the $\sharp$P-hardness of sampling statistics follows from the $\sharp$P-completeness of computing the hafnian of a random Gaussian matrix \cite{Aaronson2013,Hamilton2019}.

\section{Conclusions. Unveiling $\sharp$P-hardness \\of hybrid boson sampling statistics} 

We show that the proposed experiments on photon-atom sampling from the BEC gas of atoms and photons trapped in a multi-mode cavity have a potential to reveal $\sharp$P-hardness of sampling statistics. It is suggested by the explicit result in Eq.~(\ref{pdf=Hafnian}). In particular, one can tune to a vicinity of a confocal or concentric degeneracy point of a cavity, where there are hundreds of modes with close frequencies. Such experiments are feasible within the existing quantum-gas and cavity-QED technologies. 

Yet, they are more challenging than recent experiments \cite{Ritsch2021,Kirton2019,Mekhov2012,Esslinger2013} on phase transitions in a similar system targeted mean-field and correlation properties rather than a full quantum many-body statistics and quantum advantage. 

The hybrid boson sampler is not a quantum simulator of some input signal or controlled process. The BEC-gas in a QED cavity equipped with photon/atom detectors is just a quantum generator of random strings of photon and excited atom numbers based on a natural process of persistent quasi-equilibrium fluctuations. It is described by the statistical operator that intrinsically involves properties $\sharp$P-hard for computing. Importantly, there is no need in any controllable unitary-evolution processes (typical for quantum-computing experiments) and total suppression of relaxation and decoherence. For pioneering experiments, one should not target control of squeezing and unitary mixing (like those in Eqs.~(\ref{BlochMessiah}), (\ref{Bblocks})) in a full range aiming appearance of a truly random Gaussian block in the covariance matrix. A proof-of-principal observation of a-few-mode or two-mode squeezing and interference in the sampling statistics, showing a hafnian-like behavior as in (\ref{pdf=Hafnian}) and \cite{Entropy2023}, would be a major leap.

{}

\end{document}